\documentclass[nofootinbib,twocolumn,floatfix,preprintnumbers,superscriptaddress, longbibliography]{revtex4-1}
\usepackage[colorlinks, linkcolor=red, anchorcolor=blue, citecolor=blue, colorlinks=true, urlcolor=blue]{hyperref}
\usepackage{amssymb}
\usepackage{amsmath}
\usepackage{amsthm}
\usepackage{mathrsfs}
\usepackage{graphicx}
\usepackage{indentfirst}
\usepackage{subfigure}
\usepackage{ulem}
\usepackage{bm}
\usepackage{pifont}
\usepackage{threeparttable}
\usepackage{bbding}
\usepackage{gensymb}
\usepackage{epstopdf}
\usepackage{breakurl}
\usepackage{lipsum}
\usepackage[figuresright]{rotating}
\usepackage{fontawesome}

\begin{document}
\title{Axion effects on gamma-ray spectral irregularities with AGN redshift uncertainty}

\preprint{BNU-22-129}    

\author{Hai-Jun Li}
\email{lihaijun@bnu.edu.cn}
\affiliation{Center for Advanced Quantum Studies, Department of Physics, Beijing Normal University, Beijing 100875, China}

\author{Wei Chao}
\email{chaowei@bnu.edu.cn}
\affiliation{Center for Advanced Quantum Studies, Department of Physics, Beijing Normal University, Beijing 100875, China}

\date{\today}

\begin{abstract}

We investigate the photon-axionlike particle (ALP) oscillation effects on TeV gamma-ray spectral irregularities from the uncertain redshift active galactic nuclei (AGN) VER~J0521+211.
The gamma-ray spectra are measured by Fermi-LAT and VERITAS with the three flux states in 2013 and 2014.
We set the combined constraints on the ALP parameter ($m_a, g_{a\gamma}$) space with these states and test the extragalactic background light (EBL) absorption effect on ALP constraints with the redshift limit scenarios $z_0\sim\mathcal{O}(0.1-0.3)$.
The 99\% $\rm C.L.$ photon-ALP combined constraints set by VER~J0521+211 are roughly at $g_{a\gamma} \gtrsim 2.0\times 10^{-11} \rm \, GeV^{-1}$ for $1.0\times10^{-9} \, {\rm eV} \lesssim m_a \lesssim 1.0\times10^{-7} \, {\rm eV}$.
We find no clear connection between the redshift limit scenarios and the photon-ALP constraints.
Both the underestimated and overestimated redshifts can affect the constraint results.


\end{abstract}
\maketitle

\section{Introduction}

Axions \cite{Peccei:1977ur, Peccei:1977hh, Weinberg:1977ma, Wilczek:1977pj} and axionlike particles (ALPs) \cite{Arvanitaki:2009fg, Svrcek:2006yi} are ultralight pseudo-Nambu-Goldstone bosons (pNGBs), which are potential dark matter (DM) candidates if nonthermally generated in the early Universe through the misalignment mechanism \cite{Preskill:1982cy, Abbott:1982af, Dine:1982ah, Sikivie:2009fv, Marsh:2015xka, Chao:2022blc}. 
See $\rm e.g.$ Refs.~\cite{DiLuzio:2020wdo, Choi:2020rgn, Galanti:2022ijh} for recent reviews.
The interaction between ALPs and very high energy (VHE; $\sim \mathcal{O}(100)\, \rm GeV$) photons in the astrophysical magnetic fields with Lagrangian $-\frac{1}{4}g_{a\gamma}aF_{\mu\nu}\tilde{F}^{\mu\nu}$ could lead to detectable effects, such as a reduced TeV opacity of the Universe \cite{Mirizzi:2007hr, Simet:2007sa}.

The TeV gamma-rays from extragalactic sources are affected by the extragalactic background light (EBL) absorption effect through the pair production process, $\gamma_{\rm TeV} + \gamma_{\rm EBL} \to e^+ + e^-$.
In this case, the photon-ALP interaction provides a natural mechanism to reduce the EBL absorption and constrain the ALP properties (the ALP mass $m_a$ and the photon-ALP coupling $g_{a\gamma}$) \cite{Dominguez:2011xy, Belikov:2010ma, Horns:2012kw, DeAngelis:2011id, Meyer:2013pny, Abramowski:2013oea, TheFermi-LAT:2016zue, Guo:2020kiq, Li:2020pcn}.
The common mechanism is considering the photon-ALP conversions and back-conversions in the astrophysical magnetic fields.
If there is significant photon-ALP mixing, the Universe would appear to be more transparent than expected based on the pure EBL absorption.
See $\rm e.g.$ Refs.~\cite{Reynes:2021bpe, Li:2021gxs, Jacobsen:2022swa, Li:2022jgi, Mastrototaro:2022kpt, Dessert:2022yqq, Nakayama:2022jza, Eckner:2022rwf, Bessho:2022yyu, Bernal:2022xyi} for recent studies on photon-ALP conversions from the different extragalactic astrophysical sources. 

In this work, we focus our attention on the photon-ALP oscillation effects on TeV gamma-ray spectral irregularities from the uncertain redshift active galactic nuclei (AGN) VER~J0521+211.
VER~J0521+211 ($\rm RA=05^h21^m45^s$, $\rm Dec=21^\circ12'51.4''$, J2000) is classified as the intermediate frequency peaked BL Lac (IBL) object, which was first observed by the Very Energetic Radiation Imaging Telescope Array System (VERITAS) in 2009 \cite{VERITAS:2013nqb}.
Since the lack of optical emission features of the IBL object, the redshift of VER~J0521+211 is still unknown.
Many studies show the redshift limits of this source with $0.108\leqslant z_0 \leqslant0.34$ \cite{VERITAS:2013nqb, Shaw:2013pp, Paiano:2017pol}.
Recently, VERITAS reported the TeV gamma-ray observations of VER~J0521+211 in 2013 and 2014 with the Fermi-LAT and VERITAS data \cite{VERITAS:2022htr}, suggesting the redshift upper limits $z_0\leqslant0.31$.
Here we use these gamma-ray data and redshift limits of VER~J0521+211 to investigate the photon-ALP oscillation effects on TeV gamma-ray spectral irregularities and test the EBL absorption effect on ALP constraints.
Since the latest gamma-ray data, the redshift uncertainties, and the magnetic field parameters of VER~J0521+211 are given in Ref.~\cite{VERITAS:2022htr} together, it may be a choice for us to investigate the photon-ALP oscillation effects with this source.

This paper is structured as follows.
In Sec.~\ref{sec_gamma-ray}, we describe the VHE gamma-ray data and the redshift limits of VER~J0521+211.
In Sec.~\ref{sec_setup}, we briefly introduce the ALP constraint method and the magnetic field parameters setup.
The resulting ALP constraints are shown in Sec.~\ref{sec_constraints}.
Finally, we conclude in Sec.~\ref{sec_onclusion}.
   
\section{Gamma-ray data and redshift limits}
\label{sec_gamma-ray}

In this section, we describe the TeV gamma-ray data and the redshift limits of VER~J0521+211.
In Ref.~\cite{VERITAS:2022htr}, the gamma-ray spectra of VER~J0521+211 in 2013 and 2014 are performed by the Bayesian block (BB) analysis \cite{Scargle:2012gq}, which are defined as the flux states BB1, BB2, and BB3, respectively.
\begin{itemize}
\item  BB1: MJD 56580.0 $-$ MJD 56628.5\\
Corresponds to the intermediate state. 
\item  BB2: MJD 56628.5 $-$ MJD 56632.5\\
Corresponds to the high state. 
\item  BB3: MJD 56632.5 $-$ MJD 56689.0\\
Corresponds to the low state. 
\end{itemize}
The redshift lower limits of VER~J0521+211 are not confirmed, which can be defined as the limit scenarios L1 and L2, respectively.
\begin{itemize}
\item  L1: $z_0\geqslant0.108$ from Ref.~\cite{Shaw:2013pp}.\\
Based on a weak emission feature, which however is not confirmed by Ref.~\cite{VERITAS:2013nqb}.
\item  L2: $z_0\geqslant0.18$ from Ref.~\cite{Paiano:2017pol}.\\
The result is also not confirmed, which therefore is still unknown.
\end{itemize}
The redshift upper limit of VER~J0521+211 can be defined as the limit scenario H1.
\begin{itemize}
\item  H1: $z_0\leqslant0.308$ from Ref.~\cite{VERITAS:2022htr}.\\
The results in the literature are around $z_0\leqslant0.31$, here we take a typical value.
\end{itemize}
Using these flux states and redshift limits, we could investigate the photon-ALP oscillation effects on gamma-ray spectral irregularities.

The main effect on VHE photon (with the energy $E$) in the extragalactic space is the EBL photon (with the energy $\omega$) absorption effect with the factor $e^{-\tau}$.  
The corresponding optical depth can be described by \cite{Franceschini:2008tp}
\begin{eqnarray}
\tau=c \int_0^{z_0} \frac{{\rm d}z}{(1+z)H(z)}\int_{E_{\rm th}}^{\infty}{\rm d}\omega\frac{{\rm d}n(z)}{{\rm d}\omega} \bar{\sigma}(E,\omega,z)\, ,
\end{eqnarray}
where $H(z)=H_0\sqrt{\left(1+z\right)^2\left(1+\Omega_m z\right)-z\left(2+z\right)\Omega_\Lambda}$ is the Hubble expansion rate, with the source redshift $z_0$, the threshold energy $E_{\rm th}$, the integral pair-production cross section $\bar{\sigma}(E,\omega,z)$, the EBL proper number density ${\rm d}n(z)/{\rm d}\omega$, $H_0 \simeq 67.4\, \rm km\, s^{-1} \,Mpc^{-1}$, $\Omega_m \simeq 0.315$, and $\Omega_\Lambda \simeq 0.685$ \cite{ParticleDataGroup:2020ssz}.
In this work, the spectrum of EBL is taken from the model F-08 \cite{Franceschini:2008tp}.

In our analysis, the gamma-ray intrinsic spectrum $\Phi_{\rm int}(E)$ is selected with the minimum best-fit reduced $\chi^2_{\rm null}$ from the four spectra models as discussed in Ref.~\cite{Li:2021zms}.
Here we adopt the log-parabola model, which can be described by
\begin{eqnarray}
\Phi_{\rm int}(E)= N_0\left(\frac{E}{E_0}\right)^{-\Gamma-b \log\left(\frac{E}{E_0}\right)}\, ,
\label{eq_dnde}
\end{eqnarray}
where $N_0$ is the normalization constant, $\Gamma$ is the spectral index, $E_0$ and $b$ are free parameters. 
Then the $\chi^2$ value is given by
\begin{eqnarray}
\chi^2 = \sum_{i=1}^{N} \left(\frac{\Phi_i - \psi_i}{\delta_i}\right)^2\, ,
\label{eq_chi2}
\end{eqnarray}
with the expected spectrum $\Phi_i=e^{-\tau}\Phi_{\rm int}(E_i)$, where $N$ is the gamma-ray spectral point number, $\psi_i$ and $\delta_i$ are the detected flux and its uncertainty, respectively.

\begin{figure}[!htbp]
\centering
  \includegraphics[width=0.5\textwidth]{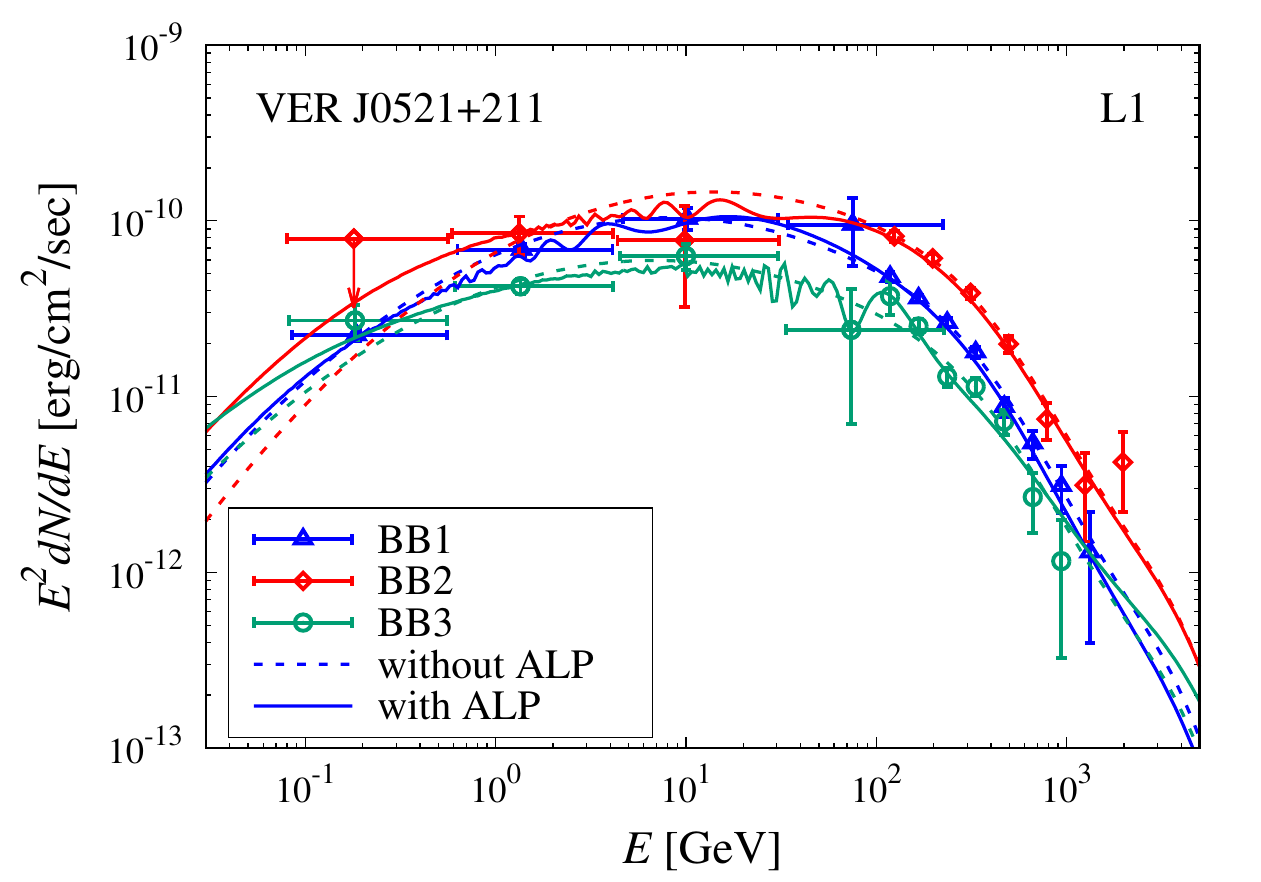} 
  \includegraphics[width=0.5\textwidth]{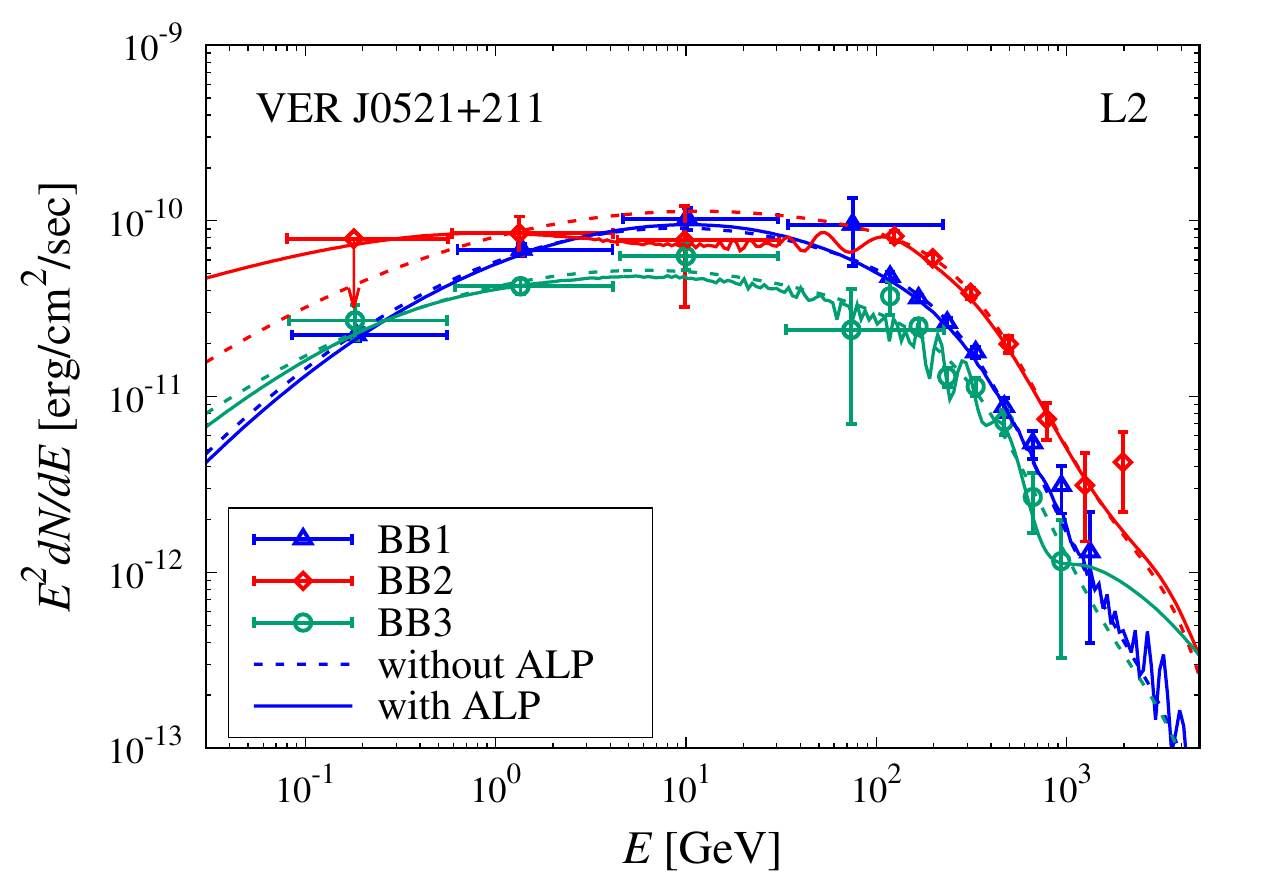} 
  \includegraphics[width=0.5\textwidth]{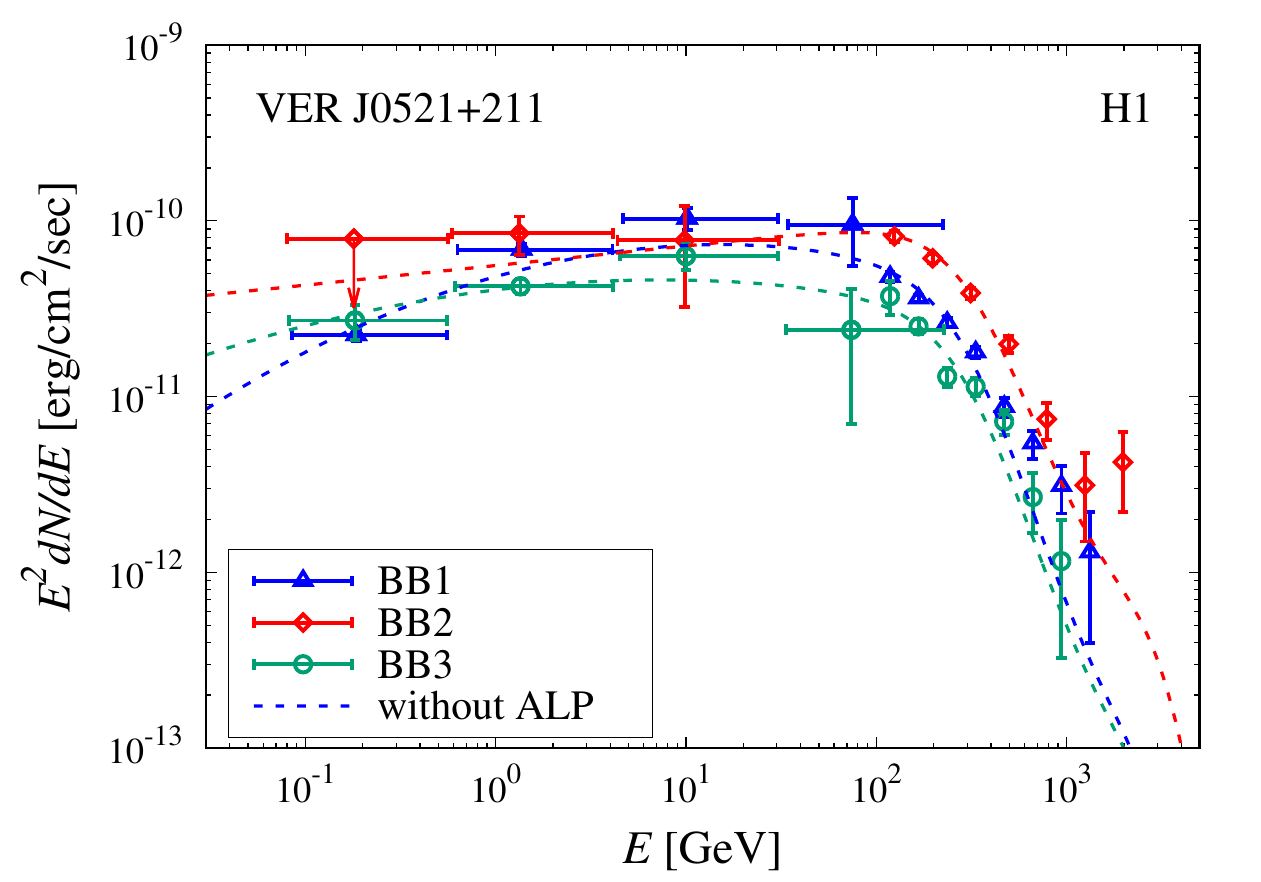}  
  \caption{The SEDs of the three states of VER~J0521+211 for the redshift limit scenarios L1 (top; with $z_0=0.108$), L2 (middle; with $z_0=0.18$), and H1 (bottom; with $z_0=0.308$). The blue triangles, red diamonds, and green circles correspond to the three states BB1 (12 points), BB2 (10 points), and BB3 (11 points), respectively. The solid and dashed lines correspond to the best-fit SEDs with/without ALP, respectively. The experimental points are taken from Fermi-LAT and VERITAS \cite{VERITAS:2022htr}. For comparison, the EBL models used in these three plots are taken as F-08 \cite{Franceschini:2008tp}.
Note that we do not make the ALP analysis with H1.  
  }
  \label{fig_dnde}
\end{figure}

\begin{table}[t]
\caption{The best-fit $\chi^2$ values under the null/ALP hypotheses of the three states of VER~J0521+211 for the scenarios L1, L2, H1, and R1. The combined results are also shown. The values of $\chi^2_{\rm min}$ correspond to the minimum best-fit points on the ALP parameter space. The effective {\rm d.o.f.} for the ALP combined analyses are also listed.}
\begin{ruledtabular}
\begin{tabular}{lcccc}
State & $\chi^2_{\rm null}$ & $\chi^2_{\rm null}/{\rm d.o.f.}$ & $\chi^2_{\rm min}$  & eff.~{\rm d.o.f.} \\
\hline
L1-BB1   &  4.89  &  0.61   &  2.43  & ...\\
L1-BB2   &  6.69  &  1.11   &  4.25  & ...\\
L1-BB3   &  14.20 &  2.03  &  8.24  & ...\\
L1-combined   &  25.78   & ...    &  18.82   & 5.58\\
\hline
L2-BB1   & 6.62   &  0.83   & 2.25   & ...\\
L2-BB2   & 3.24   &  0.54   & 2.22   & ...\\
L2-BB3   & 11.47  &  1.64   &  6.76  & ...\\
L2-combined   &  21.33   & ...    &  17.30   & 5.46\\
\hline
H1-BB1   & \boxed{44.12}  &  \boxed{5.52}  & ...  & ...\\
H1-BB2   & 15.67  &  2.61  & ...  & ...\\
H1-BB3   & 19.64  &  2.81  & ...  & ...\\
\hline
R1-BB1   &  11.92 &  1.49    &  2.50   & ...\\
R1-BB2   &  4.00   &  0.67    &  2.12   & ...\\
R1-BB3   &  12.43 &  1.78    &  7.27   & ...\\
R1-combined   & 28.35    & ...    & 18.04   &  5.41\\
\end{tabular}
\end{ruledtabular}
\label{tab_chi2}
\end{table} 

We first take the three redshift limit scenarios L1 [$z_0\sim\mathcal{O}(0.1)$], L2 [$z_0\sim\mathcal{O}(0.2)$], and H1 [$z_0\sim\mathcal{O}(0.3)$] for comparisons. 
We show the best-fit gamma-ray spectral energy distributions (SEDs) of the three states BB1, BB2, and BB3 of VER~J0521+211 with these redshift limit scenarios in Fig.~\ref{fig_dnde}.
The dashed lines represent the best-fit SEDs under the null hypothesis.
The corresponding best-fit $\chi^2_{\rm null}$ values are listed in Table~\ref{tab_chi2}.
For the scenario H1, we note that the value of $\chi^2_{\rm null}/{\rm d.o.f.}=5.52$ of BB1 is obviously larger than that of BB2 and BB3, which may be caused by the small uncertainty of the observed spectrum.
In this case ($z_0=0.308$), the other intrinsic spectra models are also checked, of which the log-parabola is still the best-fit model.
Therefore, the redshift upper limit $z_0\leqslant0.308$ may be an overestimate for BB1.
Since the large value of $\chi^2_{\rm null}/{\rm d.o.f.}$ in H1-BB1, we will not set the ALP constraint with the scenario H1.
Adopting the redshift upper limit will lead to excessive EBL absorption, and this redshift limit is not suitable to make the further analysis with other flux states (BB2 and BB3).   

\begin{figure}[!htbp]
\centering
\includegraphics[width=0.5\textwidth]{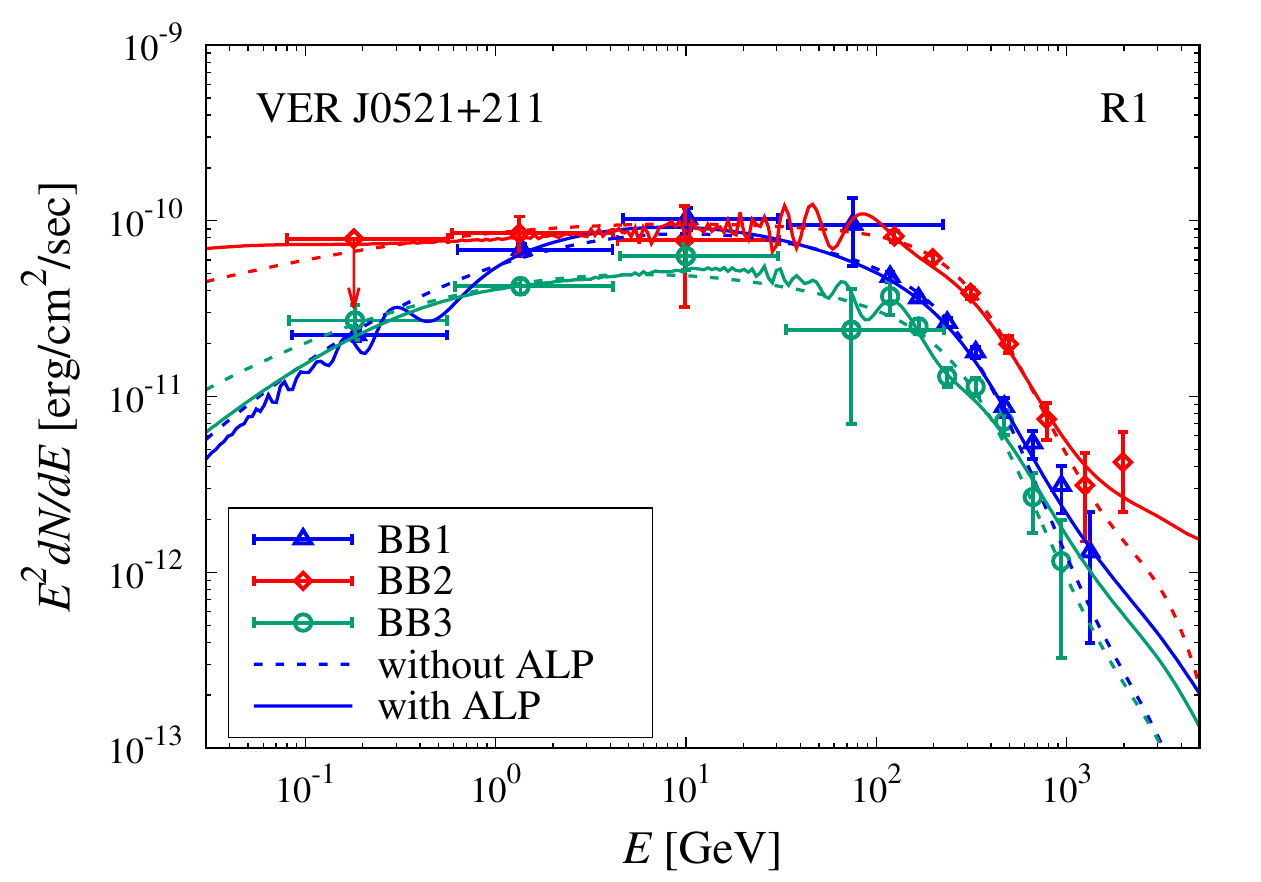} 
\caption{Same as Fig.~\ref{fig_dnde} but for the scenario R1 ($z_0=0.22$).  
}
  \label{fig_dnde_r1}
\end{figure}

In order to discuss the redshift uncertainty, here we take the another redshift value of VER~J0521+211, which can be defined as the redshift scenario R1.
\begin{itemize}
\item  R1: $z_0=0.22$.\\
This redshift value should be around at the middle of L2 and H1 ($\rm i.e.$, $z_0\simeq0.24$), and the reduced $\chi^2_{\rm null}$ of R1-BB1 should be small.
We test three values of $z_0=0.24$, 0.23, and 0.22 with R1-BB1, showing the values of $\chi^2_{\rm null}/{\rm d.o.f.}=2.12$, 1.79, and 1.49, respectively. 
Therefore, we take the value $z_0=0.22$ as R1.
\end{itemize}
For the scenario R1, we show the best-fit SEDs under the null hypothesis in Fig.~\ref{fig_dnde_r1}.
The best-fit $\chi^2_{\rm null}$ values are also listed in Table~\ref{tab_chi2}.
Compared with H1, the $\chi^2_{\rm null}$ values of R1-BB2 and R1-BB3 are also dramatically depressed.
In the following, we will just discuss the ALP hypothesis with the scenarios L1, L2, and R1.

\section{ALP setup}
\label{sec_setup}

In this section, we briefly introduce the ALP constraint method and the magnetic field parameters setup.
The photon-ALP oscillation probability in the homogeneous magnetic field can be simply described by
\begin{eqnarray}
\mathcal{P}_{a\gamma}=\left( \frac{g_{a\gamma}B_T}{\Delta_{\rm osc}}\right)^2 \sin^2\left(\frac{\Delta_{\rm osc} x_3}{2}\right)\, ,
\end{eqnarray}  
where $g_{a\gamma}$ is the photon-ALP coupling constant, $B_T$ is the transverse magnetic field, $\Delta_{\rm osc}$ is the oscillation wave number, and $x_3$ is the propagation direction of photon-ALP.
The general photon-ALP oscillations in the magnetic field can be found in Ref.~\cite{Li:2022mcf}.
Here we introduce the parameters associated the photon-ALP beam propagating from the gamma-ray source region to the Earth, which is composed of (i) the source region, (ii) the extragalactic space, and (iii) the Milky Way.
In this case, the final photon-ALP-photon oscillation probability $\mathcal{P}_{\gamma\gamma}$ for the propagation distance $s$ is given by \cite{DeAngelis:2011id}
\begin{eqnarray}
\mathcal{P}_{\gamma\gamma}={\rm Tr}\left(\left(\rho_{11}+\rho_{22}\right)\mathcal{T}(s)\rho(0)\mathcal{T}^\dagger(s)\right)\, ,
\label{eq_gammagamma}
\end{eqnarray} 
where $\mathcal{T}(s)=\mathcal{T}(s_3)_{\rm iii}\times\mathcal{T}(s_2)_{\rm ii}\times\mathcal{T}(s_1)_{\rm i}$ is the whole transfer matrix, $\rho_{11}={\rm diag}(1,0,0)$, $\rho_{22}={\rm diag}(0,1,0)$, $\rho(0)$ and $\rho(s)$ are initial and final density matrices of the photon-ALP beam, respectively. 

\begin{figure*}[t]
\centering
\subfigcapskip=0pt
\subfigbottomskip=0pt
\subfigure[~L1-BB1.]{\includegraphics[width=4.5cm]{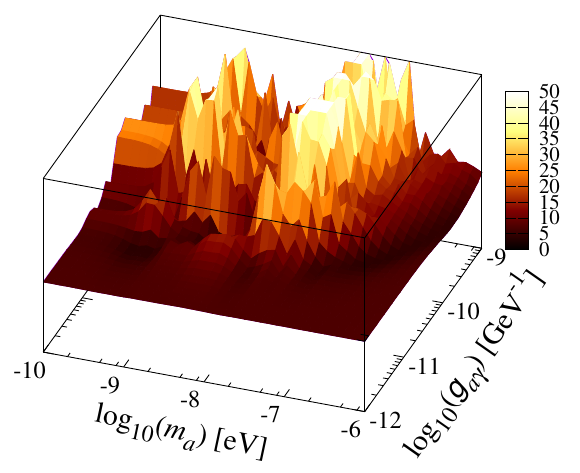}}\subfigure[~L1-BB2.]{\includegraphics[width=4.5cm]{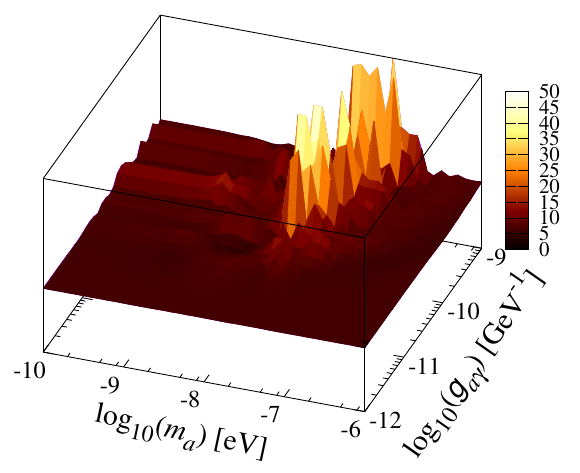}}\subfigure[~L1-BB3.]{\includegraphics[width=4.5cm]{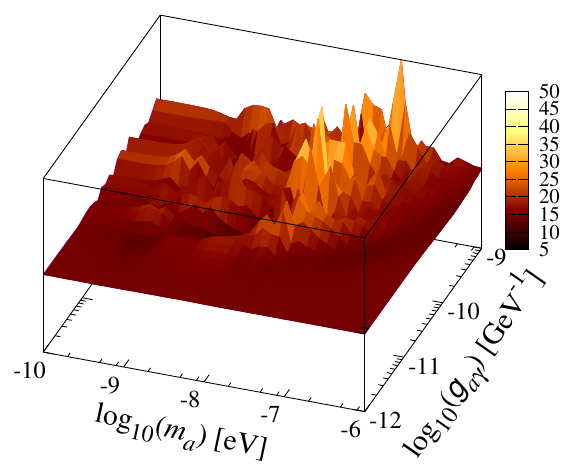}}\subfigure[~L1-combined.]{\includegraphics[width=4.5cm]{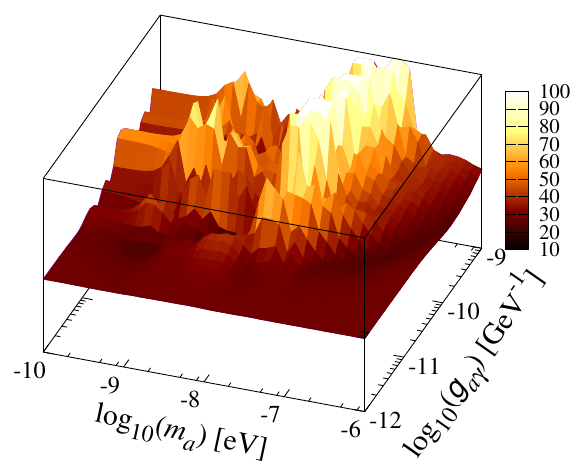}}
\subfigure[~L2-BB1.]{\includegraphics[width=4.5cm]{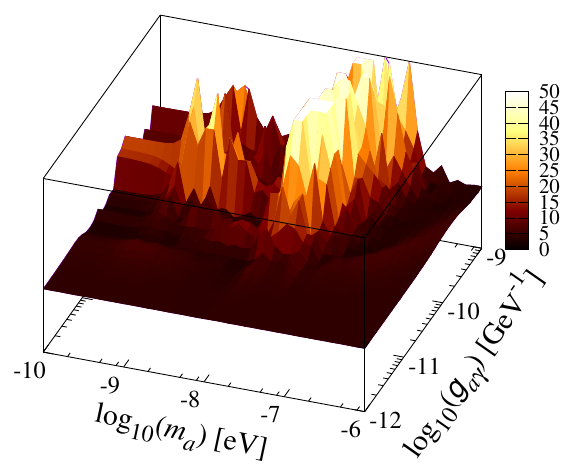}}\subfigure[~L2-BB2.]{\includegraphics[width=4.5cm]{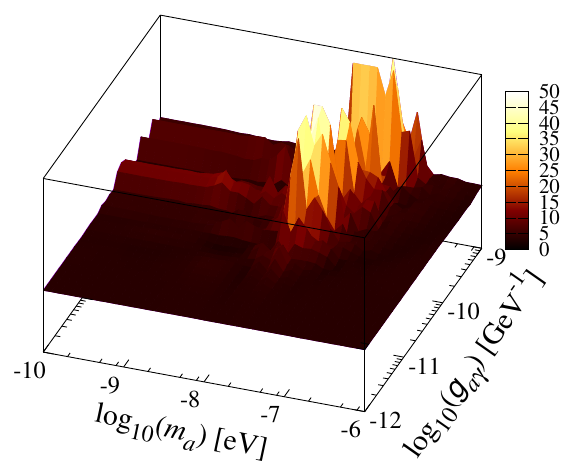}}\subfigure[~L2-BB3.]{\includegraphics[width=4.5cm]{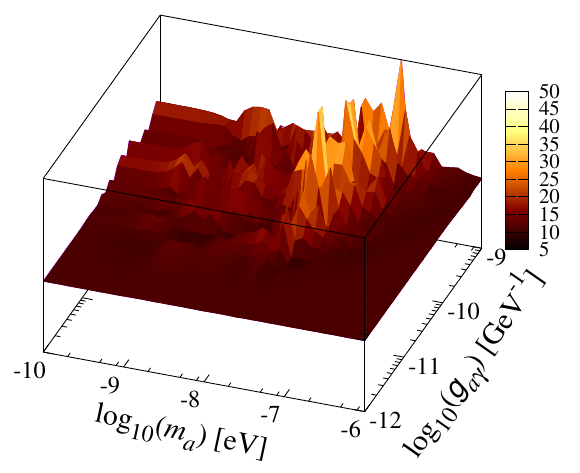}}\subfigure[~L2-combined.]{\includegraphics[width=4.5cm]{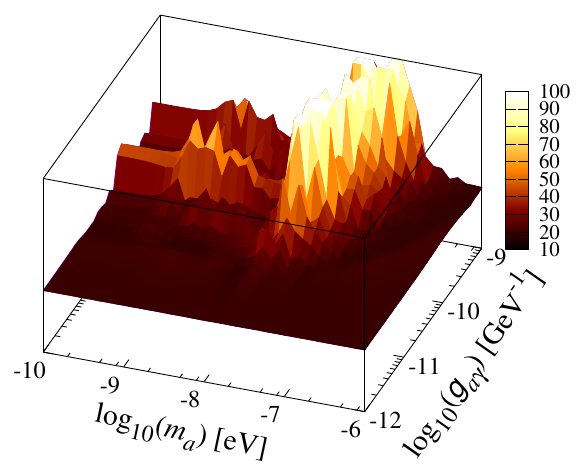}}
\subfigure[~R1-BB1.]{\includegraphics[width=4.5cm]{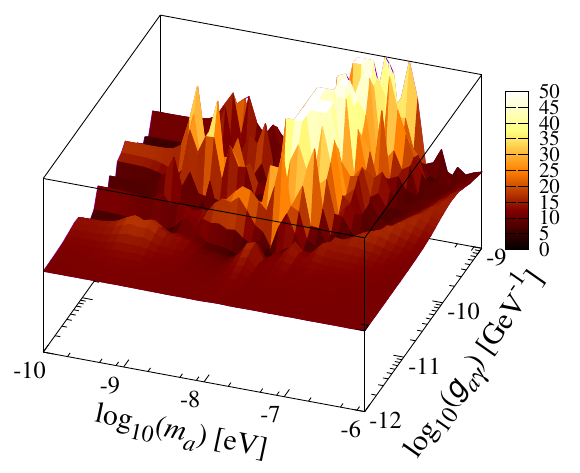}}\subfigure[~R1-BB2.]{\includegraphics[width=4.5cm]{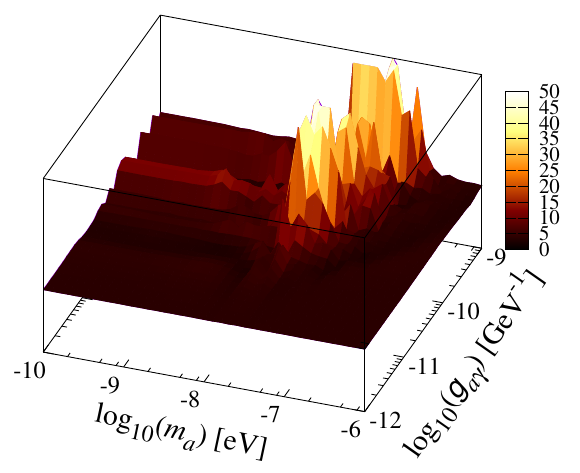}}\subfigure[~R1-BB3.]{\includegraphics[width=4.5cm]{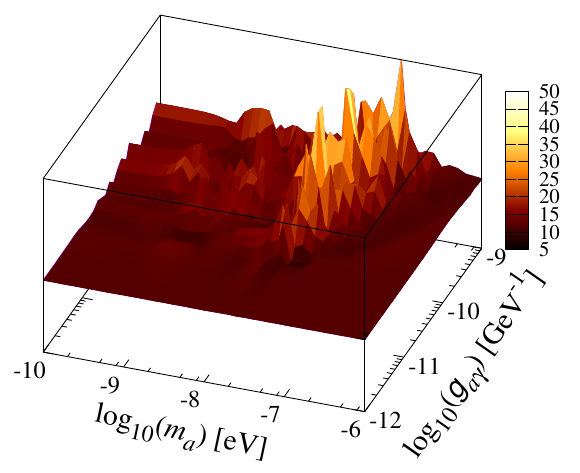}}\subfigure[~R1-combined.]{\includegraphics[width=4.5cm]{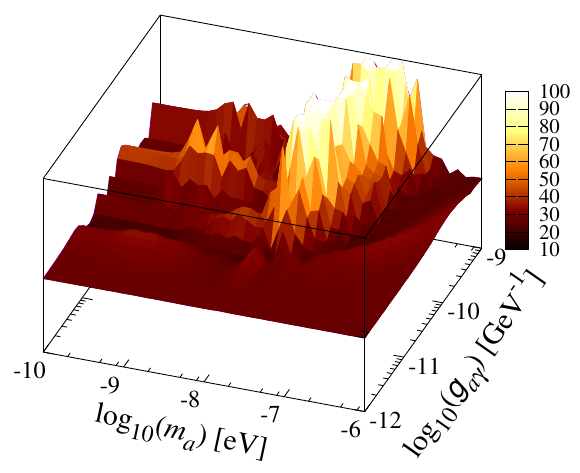}}
\caption{The best-fit $\chi^2_{\rm ALP}$ distributions on the ALP parameter ($m_a, g_{a\gamma}$) space for the redshift limit scenarios L1 (top), L2 (middle), and R1 (bottom). 
The combined results are also shown.
These panels correspond to the scenarios L1-BB1 (a), L1-BB2 (b), L1-BB3 (c), L1-combined (d), L2-BB1 (e), L2-BB2 (f), L2-BB3 (g), L2-combined (h), R1-BB1 (i), R1-BB2 (j), R1-BB3 (k), and R1-combined (l), respectively.}
\label{fig_3d}
\end{figure*}

For (i) the source region of the BL Lac object VER~J0521+211, we consider the photon-ALP oscillation in the blazar jet magnetic field, which can be described by the poloidal and toroidal components.
As discussed in Ref.~\cite{Li:2020pcn}, here we consider the jet magnetic field with the transverse magnetic field model $B(r) = B_0(r/r_{\rm VHE})^{-1}$ and the electron density model $n_{\rm el}(r) = n_0(r/r_{\rm VHE})^{-2}$, where $r_{\rm VHE}$ is the distance between the source central black hole and the VHE emission region, $B_0$ and $n_0$ are the core magnetic field and electron density at $r_{\rm VHE}$, respectively.
For the jet region $r > 1\rm\, kpc$, we take the magnetic field $B=0$.
The Doppler factor $\delta_{\rm D}=E_L/E_j$ is also considered, which represents the energy transformation between the laboratory and co-moving frames, $E_L$ and $E_j$, respectively.
The magnetic field parameters $B_0$, $r_{\rm VHE}$, $n_0$, and $\delta_{\rm D}$ for the three states BB1, BB2, and BB3 of VER~J0521+211 are listed in Table~\ref{tab_jet}.
We note that the parameters $B_0$ and $r_{\rm VHE}$ in Ref.~\cite{VERITAS:2022htr} are $1.5\times10^{-2}\, \rm G$ and $\sim\mathcal{O}(0.9)\, \rm pc$, respectively, while they are taken as $0.25\times10^{-2}\, \rm G$ and $\sim\mathcal{O}(2.5)\, \rm pc$ in Ref.~\cite{VERITAS:2013nqb}.
We also note that the latter observations are performed to constrain the ALP in Ref.~\cite{Jacobsen:2022swa} with other parameter values.
For self-consistency, the parameters setup used in this work are taken from Ref.~\cite{VERITAS:2022htr}.
Additionally, for the host galaxy region of VER~J0521+211, the photon-ALP oscillation in this part can be totally neglected.

For (ii) the extragalactic space, we just consider the EBL absorption effect on VHE photon due to the pair-production process.
Since the magnetic field in the extragalactic space is very weak with the upper limit $\sim\mathcal{O}(1)\, \rm nG$ \cite{Ade:2015cva, Pshirkov:2015tua}, we neglect the photon-ALP oscillation in this region.

Finally, we also take into account the photon-ALP oscillation in (iii) the Milky Way with the Galactic magnetic field model \cite{Jansson:2012pc, Jansson:2012rt}, which is composed of the disk and halo components (both parallel to the plane of the Milky Way), and the so-called ``X-field" component (out-of-plane) at the center of the Milky Way.
See also Refs.~\cite{Planck:2016gdp, Unger:2017kfh} for the latest version of this model.

\begin{table}[t]
\caption{The source jet magnetic field parameters of the three states of VER~J0521+211. These values can be directly or indirectly obtained from Ref.~\cite{VERITAS:2022htr}.}
\begin{ruledtabular}
\begin{tabular}{lcccc}
State & $B_0$($10^{-2}\, \rm G$) & $r_{\rm VHE}$(pc)  &$n_0$($10^{3} \, \rm cm^{-3}$)& $\delta_{\rm D}$  \\
\hline
BB1   & 1.5  &  0.85  & 0.88 & 26 \\
BB2   & 1.5  &  0.89  & 0.95 & 26 \\
BB3   & 1.5  &  0.93  & 0.68 & 26 \\
\end{tabular}
\end{ruledtabular}
\label{tab_jet}
\end{table}

\section{ALP constraints with the redshift uncertainty}
\label{sec_constraints}

Using Eqs.~(\ref{eq_dnde}) and (\ref{eq_gammagamma}), the expected gamma-ray spectrum under the ALP hypothesis can be described by $\Phi_{{\rm ALP},\, i} = \mathcal{P}_{\gamma\gamma} \Phi_{\rm int}(E_i)$, where $\Phi_{\rm int}(E_i)$ is the gamma-ray intrinsic spectrum.
For one ALP parameter ($m_a, g_{a\gamma}$) set, we can derive the best-fit $\chi^2_{\rm ALP}$ from Eq.~(\ref{eq_chi2}) with the notation $\Phi_i\to\Phi_{{\rm ALP},\, i}$.
Then we can derive the best-fit $\chi^2_{\rm ALP}$ distributions on the whole ALP parameter space, which are shown in Fig.~\ref{fig_3d} with the redshift limit scenarios L1, L2, and R1.
In these panels, the distributions of $\chi^2_{\rm ALP}$ correspond to the three states BB1, BB2, and BB3 of VER~J0521+211.
The minimum best-fit gamma-ray SEDs under the ALP hypothesis of these three states with the scenarios L1, L2, and R1 are also shown in Fig.~\ref{fig_dnde} and \ref{fig_dnde_r1} for comparisons, respectively.
The values of minimum best-fit $\chi^2_{\rm min}$ on the ALP parameter space can be found in Table~\ref{tab_chi2}.
Compared with the null hypothesis, the minimum best-fit $\chi^2_{\rm min}$ under the ALP hypothesis can be dramatically depressed.

As considered in Refs.~\cite{Li:2020pcn, Li:2021gxs}, we also set the combined constraints on ALP with the multistate analysis.
In this case, the two or more states of the same source are selected to fit with the corresponding magnetic field setup.
In order to obtain the $\chi^2_{99\%}$ value at 99\% $\rm C.L.$, 400 sets of the gamma-ray spectra observations in the pseudoexperiments by Gaussian samplings are simulated to derive the test statistic (TS) distribution, ${\rm TS}={\widehat{\chi}_{\rm null}}^2 - {\widehat{\chi}_{\rm ALP}}^2$, with the best-fit $\chi^2$ of the null and ALP hypotheses in the Monte Carlo simulations, ${\widehat{\chi}_{\rm null}}^2$ and ${\widehat{\chi}_{\rm ALP}}^2$, respectively.
Here the TS distribution obeys the non-central $\chi^2$ distribution with the effective ${\rm d.o.f.}$ and the non-centrality $\lambda$.
Then we assume this TS distribution is approximated with the ALP hypothesis and can be used to derive $\Delta\chi^2_{99\%}$.
Finally, the value of the 99\% $\rm C.L.$ $\chi^2$ can be obtained by $\chi^2_{99\%}=\chi^2_{\rm min}+\Delta\chi^2_{99\%}$.
More details about the statistical method can be found in Ref.~\cite{Li:2022jgi}.

\begin{figure}[tbp]
\centering
\includegraphics[width=0.49\textwidth]{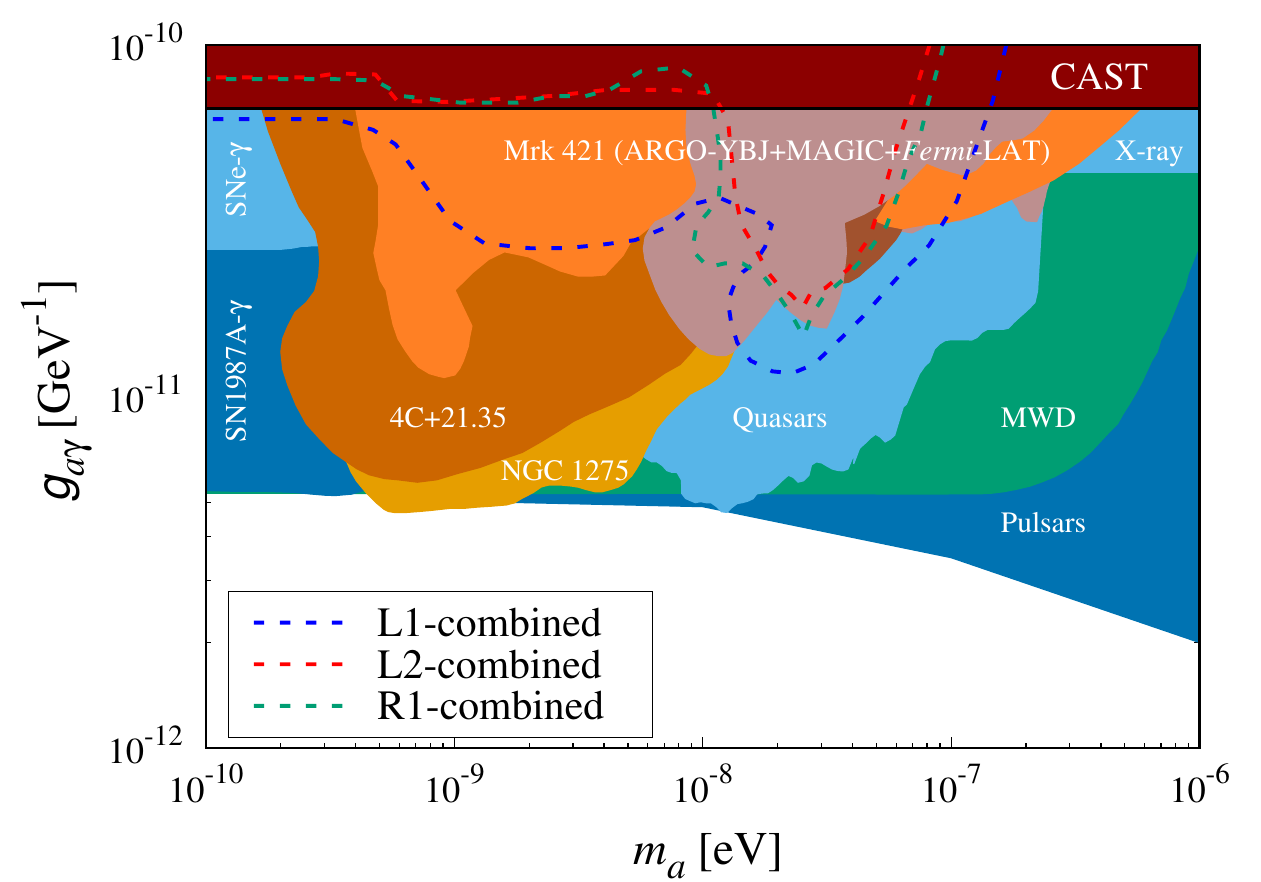}  
\caption{The 99\% $\rm C.L.$ photon-ALP combined constraints set by VER~J0521+211. The dashed blue, red, and green contours represent the 99\% $\rm C.L.$ combined results of the scenarios L1-, L2-, and R1-combined, respectively. The other limits are taken from the package {\tt AxionLimits} \cite{ciaran_o_hare_2020_3932430}.}
  \label{fig_compare}
\end{figure}

Then we show the ALP combined constraint results set by VER~J0521+211 in Fig.~\ref{fig_compare}.
For comparison, we also show the other latest photon-ALP constraints \cite{ciaran_o_hare_2020_3932430} in this plot.
The dashed blue, red, and green contours represent the 99\% $\rm C.L.$ combined results of L1, L2, and R1 with the three states combined, respectively.
The corresponding best-fit $\chi^2_{\rm min}$ values and the effective ${\rm d.o.f.}$ are also listed in Table~\ref{tab_chi2} with $\lambda=0.01$.
Compared with the two redshift limit scenarios L1 ($z_0\geqslant0.108$) and L2 ($z_0\geqslant0.18$), we find the more stringent ALP combined constraint with the underestimated redshift. 
The 99\% $\rm C.L.$ exclusion region of L2-combined is completely covered by L1-combined, which shows significant difference in the low mass region $1.0\times10^{-9} \, {\rm eV} \lesssim m_a \lesssim 1.0\times10^{-8} \, {\rm eV}$.
This is probably because the null hypothesis SEDs of L1-BB2 and L1-BB3 in low energies ($0.1-1\, \rm GeV$) cannot be fitted well with $z_0=0.108$, see Fig.~\ref{fig_dnde}.
While compared with the scenarios L2 and R1 ($z_0=0.22$), we find they show similar ALP exclusion regions (the dashed red and green contours in Fig.~\ref{fig_compare}).
The small difference of redshifts (0.18 and 0.22) will reduce the constraint difference on the ALP parameter space.
Therefore, no clear connection is confirmed between the redshift limit scenarios and the photon-ALP constraints.

Finally, we give the 99\% $\rm C.L.$ ALP combined constraints set by VER~J0521+211, which are roughly at $g_{a\gamma} \gtrsim 2.0\times 10^{-11} \rm \, GeV^{-1}$ for $1.0\times10^{-9} \, {\rm eV} \lesssim m_a \lesssim 1.0\times10^{-7} \, {\rm eV}$.
Additionally, we note that our results are generally similar to the 99\% $\rm C.L.$ limit set by the 2009 TeV observations \cite{VERITAS:2013nqb} of VER~J0521+211 in Ref.~\cite{Jacobsen:2022swa}, which is performed with the different data (VERITAS+HAWC) and magnetic field parameters.
Since HAWC can measure the VHE gamma-rays exceeding $\sim1-100\,\rm TeV$, in this case, the photon-ALP oscillation effects will become more significant.
However, the limited energy resolution of HAWC can also affect the ALP signal sensitivity in this high energy region.

\section{Conclusion}
\label{sec_onclusion}

In this paper, we have presented the effects of photon-ALP oscillation on TeV gamma-ray spectral irregularities from the AGN VER~J0521+211, which is classified as the IBL object with the uncertain redshift $0.108\leqslant z_0 \leqslant0.34$.
The gamma-ray spectra are measured by Fermi-LAT and VERITAS in 2013 and 2014 with the three flux states (BB1, BB2, and BB3), and analyzed with the four redshift limit scenarios L1 ($z_0\geqslant0.108$), L2 ($z_0\geqslant0.18$), H1 ($z_0\leqslant0.308$), and R1 ($z_0=0.22$).
The SEDs of these states under the null and ALP hypotheses are shown for comparisons.
Then we set the combined constraints on the ALP parameter space with these states and test the effect of EBL absorption on ALP constraints with these redshift limit scenarios.
Since the redshift upper limit scenario H1 may be an overestimate for BB1, we do not set the ALP constraint with H1.
In addition, we take the another redshift scenario R1 to discuss the redshift uncertainty.

The 99\% $\rm C.L.$ photon-ALP combined constraints set by the scenarios L1, L2, and R1 on the ALP parameter space are roughly at $g_{a\gamma} \gtrsim 2.0\times 10^{-11} \rm \, GeV^{-1}$ for $1.0\times10^{-9} \, {\rm eV} \lesssim m_a \lesssim 1.0\times10^{-7} \, {\rm eV}$.
Compared with the results of L1- and L2-combined, we find the more stringent ALP combined constraint with the underestimated redshift. 
While compared with the scenarios L2-and R1-combined, we find they show similar ALP exclusion regions.
Then we have the conclusion that no clear connection is confirmed between the redshift limit scenarios and the photon-ALP constraints.
Both the underestimated and overestimated redshift limit scenarios can affect the constraint results.
In this work, the underestimated redshift (L1) shows a stringent ALP exclusion region, while the overestimated redshift (H1) is not suitable to make the ALP analysis.

\section*{Acknowledgments}
The authors would like to thank Qi Feng for sharing the experimental data of VER~J0521+211 and valuable comments, and also thank Peng-Fei Yin for helpful discussions. 
This work was supported by the National Natural Science Foundation (NNSF) of China (Grants No.~11775025 and No.~12175027).

\bibliography{references}
\end{document}